\documentclass[11]{article}

\usepackage{hyperref}	
\usepackage{amsmath}	
\usepackage{graphicx}	
\usepackage{fullpage}	
\graphicspath{{SNewpaper/}}
\usepackage{ amssymb }
\usepackage[T1]{fontenc}
\usepackage[utf8]{inputenc}
\usepackage{authblk}

\begin{document}


\title{Relations Involving Static Quadrupole Moments of $2^{+}$ states
and B(E2)'s}	

\author[*]{S. Yeager}

\author[*]{L. Zamick}

\author[*]{Y.Y. Sharon}

\author[*]{Xiaofei Yu}

\affil[*]{Department of Physics and Astronomy, Rutgers University Piscataway,
New Jersey 08854}

\author[+]{S.J.Q. Robinson}

\affil[+]{Department of Physics, Millsaps College Jackson, Mississippi 39210}
\maketitle


\begin{abstract}
We define the {}``quadrupole ratio'' $r_{Q}=\dfrac{Q_{0}(S)}{Q_{0}(B)}$
where $Q_{0}(S)$ is the intrinsic quadrupole moment obtained from
the static quadrupole moment of the $2_{1}^{+}$ state of an even-even
nucleus and $Q_{0}(B)$ the intrinsic quadrupole moment obtained from
$B(E2)_{0\rightarrow2}$. In both cases we assume a simple rotational
formula connecting the rotating frame to the laboratory frame. The
quantity $r_{Q}$ would be one if the rotational model were perfect
and the energy ratio $E(4)/E(2)$ would be $10/3$. In the simple
vibrational model, $r_{Q}$ would be zero and $E(4)/E(2)$ would be
two. There are some regions where the rotational limit is almost met
and fewer where the vibrational limit is also almost met. For most
cases, however, it is between these two limits, i.e. $0<|r_{Q}|<1$.
There are a few cases where $r_{Q}$ is bigger than one, especially
for light nuclei. In most cases the quadrupole ratio is positive but
there are two regions with negative ratios. The first case is that
of light nuclei and the second has certain nuclei close to $^{208}$Pb. 
\end{abstract}

\section{Introduction}

In previous works \cite{robinson06,escuderos} a certain ratio was
defined involving the static quadrupole moment of a $2^{+}$ state
in an even-even nucleus and the $B(E2)_{0\rightarrow2}$. It was defined
in such a way so that for a perfect rotor, the ratio was one and for
a simple vibrator, the ratio was zero.

In the rotational model the $Q(2^{+})$ and $B(E2)_{0\rightarrow2}$
are described by one parameter, the intrinsic quadrupole moment Q$_{0}$.
We will however define two different operational static quadrupole
moments.

\begin{subequations} 
\begin{gather}
B(E2)_{I1\rightarrow I2}=\frac{5}{16\pi}Q_{0}^{2}(B)|\langle I_{1}\;\; K\;\;2\;0\;\;\mid\;\; I_{2}\;\; K\rangle|^{2}\\
Q(I)=\frac{3K^{2}-I(I+1)}{(I+1)(2I+3)}Q_{0}(S)
\end{gather}
 \end{subequations}

Where

\begin{subequations} 
\begin{gather}
Q_{0}(B)=\sqrt{B(E2)}\sqrt{\frac{16\pi}{5}}\\
Q_{0}(S)=-\frac{7}{2}Q(I)
\end{gather}
 \end{subequations}

For $K=0$, $I_{1}=0$, $I_{2}=2$ we obtain

\begin{subequations} 
\begin{gather}
B(E2)_{0\rightarrow2}=\frac{5}{16\pi}Q_{0}^{2}(B)\\
Q(2^{+})=-\frac{2}{7}Q_{0}(S)
\end{gather}
 \end{subequations}

The quadrupole ratio is

\begin{align}
r_{Q}=\frac{Q_{0}(S)}{Q_{0}(B)} & =-\frac{7}{2}\sqrt{\frac{5}{16\pi}}\frac{Q(2^{+})}{\sqrt{B(E2)_{0\rightarrow2}}}\\
 & =-1.1038\frac{Q(2^{+})}{\sqrt{B(E2)_{0\rightarrow2}}}\nonumber 
\end{align}

Thus, we have expressed the quadrupole ratio in terms of quantities
measured in the laboratory. For $Q(2^{+})$ we use the reference of
Stone \cite{stone05} and for B(E2), the tables of Raman et al. \cite{raman01}.

The main difference from the previous works is that we now consider
all nuclei for which $Q(2^{+})$ has been measured. In previous works,
only light nuclei were considered. Looking at all nuclei, as we do
here, will give us a new perspective.

\section{Choices made in getting $Q(2^{+})$}

It is much easier to measure $B(E2)_{0\rightarrow2}$ than it is to
measure $Q(2^{+})$. The values of $B(E2)$ in Raman's table form
a concensus made by Raman when more than one measurement was made
on a nucleus.

For $Q(2^{+})$, Stone does not present an evaluated result, and wisely
so, because there is a wide variation in some cases. What was done
in this work was to select the latest measurement of a given $Q(2^{+})$.
Of course it is not always true that the latest measurement is the
best one. Hence, if later, more definate measurements are made, we
will have to alter the figures. We feel that the large scale impression
of the results will still stand.

Note that Stone's table does not contain measured values of $Q(2_{1}^{+})$
beyond $^{206}$Pb. Hence in this analysis we cannot comment on nuclei
beyond doubly magic $^{208}$Pb such as the actinide nuclei and the
transuranic nuclei.

\section{Results}

In Figure I we present for even-even nuclei the magnitude ofthe quadrupole
ratio (with $-\dfrac{7}{2}\sqrt{\dfrac{5}{16\pi}}$ replaced by k)
and $E(4)/E(2)$, the ratio of energies. Again, in the pure rotation
model $r_{Q}$ is equal to 1 and $E(4)/E(2)=\dfrac{4\times5}{2\times3}=3.33$.
In a simple vibrational limit the quadrupole moment would be zero
(i.e. the static quadrupole moment would vanish) and $E(4)/E(2)=2$.

\begin{figure}
\begin{center}
\includegraphics[scale=0.35]{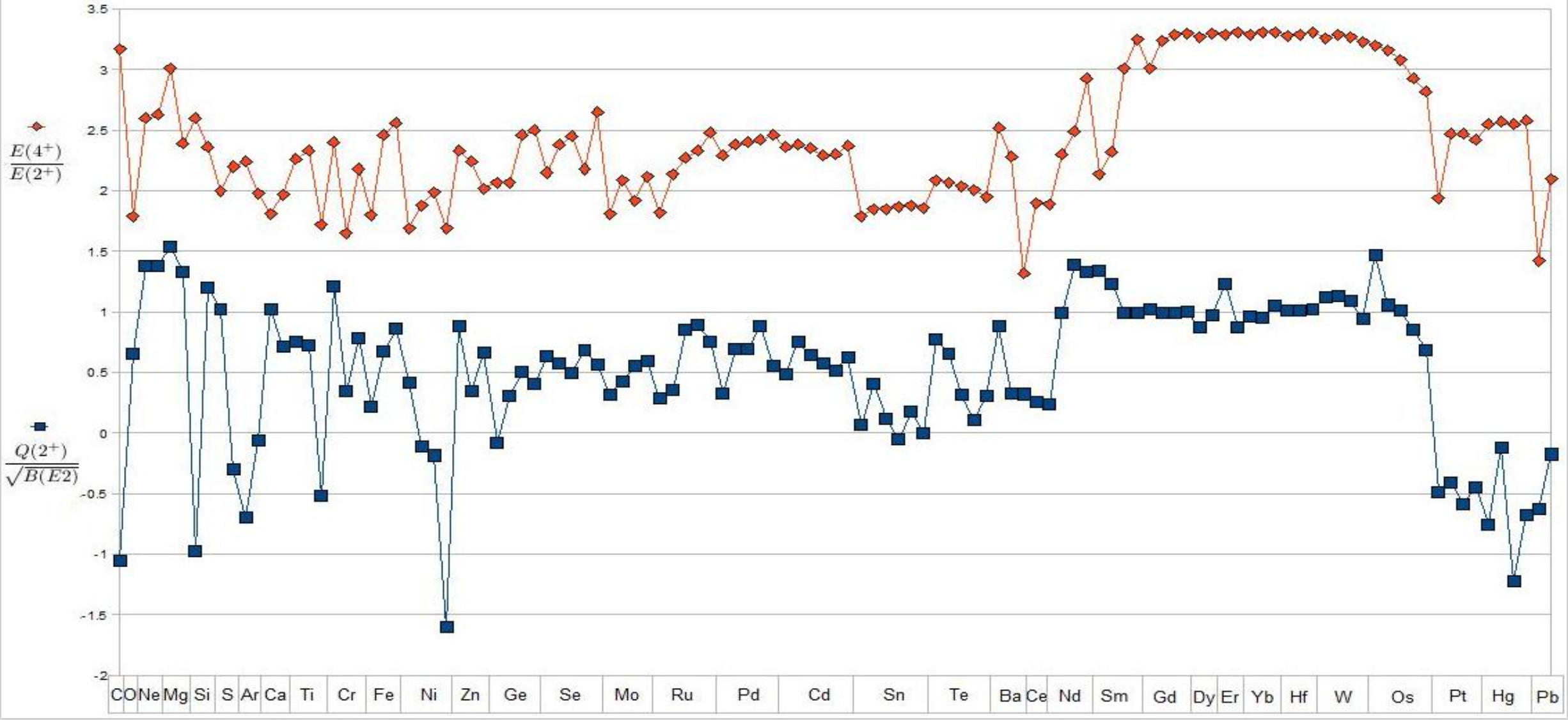}
\end{center}
\caption{Data on Static  Quadruple Moments \label{fig:1}}
\end{figure}

For light nuclei there are several examples where the quadrupole ratio
is even larger than one. There is a midrange region from about 70
to 110 where the quadrupole ratio is around 0.5 and where $E(4)/E(2)$
is between two and three. This is a region which is about halfway
between vibrational and rotational. Then we hit an island 112 and
120 where thhe simple vibrational model is almost realized. The nuclei
in this band include $^{112,118,120,122,124}$Sn and $^{128}$Te.
The ratio here for $Q(2^{+})$ is almost zero and $E(4)/E(2)$ is
close to two.

As we move to heavier nuclei, the rotational model is realized in
a band from about 152 to 186, with the quadrupole ratio close to one
and $E(4)/E(2)$ close to 10/3. These nuclei include $^{146,148,150}$Nd,
$^{152,154}$Sn, Gd isotopes, $^{160}$Dy, $^{164}$Dy, Er isotopes,
Yb isotopes, $^{182,184,and186}$W, most Os isotopes, and $^{202}$Hg.

We can also consider other models. For example we can use the $f_{7/2}$
configuration for $^{42}$Ca. This single particle model is the antithesis
of the rotational model. Yet it yields a value $r_{Q}=0.78$. This
compares with the experimental value of 1.023.

We next discuss the signs of the quadrupole ratios. For the light
nuclei there are several cases where the quadrupole ratio is negative
($^{12}$C, $^{28}$Si, $^{34}$S, $^{36}$Ar, $^{40}$Ar, ${50}$Ti,
$^{60}$Ni, $^{62}$Ni, $^{64}$ Ni, and $^{70}$Ge). It was pointed
out by Bohr and Mottelson that in the absence of a spin-orbit interaction
about half the nuclei would be prolate and the other half oblate \cite{bohr}.
It is known that the spin orbit force is not as important in the light
nuclei as it is for heavier ones, so this is consistent.

We also encounter nuclei with positive values of Q(2+) as we approach
$^{208}$Pb, namely $^{192,194,196,198}$Pt, $^{198,200,202,204}$Hg
and $^{204,206}$Pb. It is tempting to refer to these as oblate but
more complex analyses are present \cite{ciz}, especially for the
Pt isotopes.

\section{Deformation in terms of Mottelson Conditions}

In Hartree Fock codes, the deformation parameter is defined as 
\begin{equation}
\beta_{0}=\sqrt{\dfrac{\pi}{5}}\dfrac{Q_{0}(protons)}{2<r^{2}>_{protons}}
\end{equation}
 where Q$_{0}$ is the intrinsic quadrupole moment and $<r^{2}>$
is the mean square radius of the protons. In the deformed oscillator
model the Hamiltonian is H=$\hbar\omega_{x}\Sigma_{x}+\hbar\omega_{y}\Sigma_{y}+\hbar\omega_{z}\Sigma_{z}$
where $\Sigma_{x}$ is the sum of (2N$_{x}$+1/2) over all occupied
orbits.

If we have axial symmetry H=$\hbar\omega_{x}(\Sigma_{x}+\Sigma_{y})+\hbar\omega_{z}\Sigma_{z}$,
we can show that 
\begin{equation}
\sqrt{\frac{5}{\pi}}\beta_{0}=\frac{2\Sigma_{z}\omega_{x}-(\Sigma_{x}+\Sigma_{y})\omega_{z}}{(\Sigma_{x}+\Sigma_{y})\omega_{z}+\Sigma_{z}\omega_{x}}
\end{equation}

The Mottelson conditions are $\omega_{x}\Sigma_{x}=\omega_{y}\Sigma_{y}=\omega_{z}\Sigma_{z}$.
Using them we get 
\begin{equation}
\beta_{0}=\sqrt{\dfrac{\pi}{5}}\frac{2(\Sigma_{Z}^{2}-\Sigma_{x}^{2})}{2\Sigma_{x}^{2}+\Sigma_{z}^{2}}
\end{equation}

We see that we get prolate deformation if $\Sigma_{z}$ is greater
than $\Sigma_{x}$ and oblate if $\Sigma_{z}$ is less than $\Sigma_{x}$.

\section{Closing Remarks}

This work represents an expansion of the 2006 work of Robinson et
al. \cite{robinson06}. We here consider essentially all nuclei where
$Q(2^{+})$ has been measured. Since that work, there have been other
works of relevance e.g. that of Bertsch et al. \cite{bertsch07} in
which the Gogny interaction was used to calculate $Q(2^{+})$ and
$B(E2)$ as well as work by Sabbey et al. \cite{sabbey07}. Early
works using Skyrme H.F. were performed by Jaqaman et al. \cite{jaqaman84}
who also considered hexadecapole models. Recent work by Sarriguren
et al. \cite{sarriguren} should be noted as well as the phase transitions
in the platinum isotopes by Morales et al.\cite{morales08}. They
consider the heavier nuclei -- isotopes of Yb, Hf, W, Os, and Pt --
where there are many changes from prolate to oblate. Also mentioned
in the 2006 work, Zelevinsky and Volya \cite{zelevinsky04} noted
that with random matrices they obtained with a high probability that
the quadrupole ratio $r_{Q}$ (as we define it) is either one or zero.
Why this is so is not clear.

We thank the Aresty program at Rutgers for their support. We thank
Noemie Koller, Jolie Cizewski, and Gulhan Gurdal for critical but
cogent comments, and Gerfried Kumbartzki for his help.

\end{document}